# Retracing and Assessing the CEP Project

**Giovanni A. Cignoni, Fabio Gadducci**
University of Pisa

**Abstract.** The last decade witnessed a renewed interest in the development of the Italian computer industry and in the role of the Fifties pioneers in Rome, Milan, Ivrea, and Pisa. The aim of the paper is to retrace some steps of the CEP project, carried out by the University of Pisa in collaboration with Olivetti, by reassessing the documents preserved in the University archives. The project was a seminal enterprise for Italy, and among its accomplishments it delivered in 1957 the first Italian computer. The mix of public sector funding and industrial foretelling witnessed by the project is one of the leading examples in Italy of best practices, and its success paved the way for the birth of Computer Science in the country as an industry as well as a scientific discipline.

## Introduction

Despite the attention devoted early on to some of the protagonists, namely the Olivetti firm and the figures of Adriano and Roberto Olivetti,[1] the research on the history of Italian computer science started quite late, the seminal event being the 1991 conference organized by Italian Association for informatics and Automatic Computing (AICA).[2] However, the last decade has been fruitful of new investigations, the starting point being the 2004 Milan meeting celebrating the fifty years of the arrival of the first computer in Italy,[3] which were followed by events focussing on the Rome[4] and Pisa[5] accomplishments, covering the centres that introduced computer science in Italy.[6]

All these events celebrated the pioneering Italian experiences of the Fifties, usually with the participation of the protagonists of the period and often recording their personal recollections. The current research has however moved beyond such memories, providing a documentary context as well as pushing the exploration of the archives. Concerning Pisa, there are now repositories making available many original documents,[7] which are now starting to be assessed.[8] Indeed, the focus of the paper is on retracing less-explored facets of the *CEP project* (for Calcolatrice Elettronica Pisana, i.e. Pisa Electronic Calculator), carried on by the local University. The project started in 1954 with the ambitious goal of building from scratch an electronic computer, the first one of its kind in Italy. The main partner was Olivetti, which used the expertise to start its own line of commercial computers, the first one to be marketed being the transistorised ELEA 9003 in 1959.[9]

More specifically, we aim at going again over the documents held by the Archives of the University of Pisa to further explore the scenario sketched by the previous literature. We especially searched internal reports and blueprints by *Centro Studi Calcolatrici Elettroniche* (CSCE, Centre for the Study of Electronic Calculators), the department in charge of the CEP project that was established by the University. Then we assessed the documents against the technological knowledge of the period. The careful analysis of the contents, combined with the re-reading of the administrative papers as well as further investigations in other archives and personal correspondences, resulted in a new perspective on the accomplishments of the project and its impact on the by then blooming Italian computer industry.

The paper will focus on three issues: clarifying the beginning of the project and the actual involvement of Enrico Fermi; restoring the memories and the technical achievements of the very first Italian computer, built in 1957 during the CEP project; and recognizing the difficulties of the second phase of the project and their impact on the final machine delivered in 1961.

## The birth of CSCE and the role of Enrico Fermi

A firmly established part of the lore about the CEP project concerns the role of the renowned scientist Enrico Fermi in the decision to build a computer, using the funds granted to the University of Pisa by a consortium of local administrations, the *Consorzio Interprovinciale Universitario* (CIU, meaning Inter-province Consortium for the University).

It is also well known that building a computer was not the first choice. The CIU funds, provided by the municipalities and provinces of Pisa, Livorno and Lucca, should have been used for the construction of a synchrotron, which was designed by the Pisa Institute of Physics. Eventually, the synchrotron was built in Frascati using funds provided by the Rome municipality. Then, the CIU funds were used for the CEP project.

It is thus surprising that the archives preserve very few traces concerning the funding of the synchrotron. In the minutes of the Executive Committee of the CIU the synchrotron is mentioned only on May 20, 1955,[10] when this project has since long moved away and the decision to build a computer was already taken. A summary[11] cites an unrecorded meeting of March 20, 1954 during which, most likely, the commitment to the synchrotron was decided. The summary itself however is of a much later date, as it already mention the choice of the new splitting of the funds between the electronic calculator and the mass spectrograph established at the meeting of October 4, 1954.[12]

When compared to reports in the recent literature, also surprising is the actual involvement of Enrico Fermi in the choice for the new use of the CIU funding. From the surviving documents and the correspondence between the protagonists we can infer that the role of the famous scientist was different from the current tradition of a direct suggestion. Besides the letters between Fermi and Enrico Avanzi, then Rector of the University of Pisa, in August 1954[13], the facts can be inferred from the letters between Marcello Conversi, [14] director of the Institute of Physics, and Gilberto Bernardini,[15] president of the National Institute of Nuclear Physics (INFN), with Mauro Picone, head of the National Institute for Applied Mathematics (INAC) in Rome.

Fermi is often recalled[16] as single-handedly pushing the built of an electronic computer in Varenna, in the occasion of the International Physics School meeting. On the contrary, while the location is correct, the documents tell of a "discussion lasting for days", during which the possible new uses of the CIU funds were debated "in an atmosphere of dispassionate objectivity and clarity."[17] The outcome was that the construction of an electronic computer was "by far the best [choice] among all the others", as Fermi stated in an endorsement letter to the Pisa Rector.[18] Such letter was not an autonomous initiative of Fermi: as Conversi states it was written "at mine and prof. Bernardini's request".[19] Avanzi replied to Fermi saying the he was "pleased that he had discussed with the colleagues Conversi and Salvini about the possibility of equipping the University of Pisa of a scientific instrument of national importance,"[20] suggesting that the Varenna meeting had been considered by the University, and in general by the whole community of Italian physicists, as the ideal context to discuss the new destination of the funds, in order to submit to CIU a proposal that already had a strong authoritative support from a large number of scientists (not only from Pisa).

Nevertheless, the actual role played by Fermi helps in understanding the context in which the CEP project was born. Fermi was aware of the importance of computers for research and industry[21] and his endorsement was crucial to unlock the hesitations in Pisa. In October two important politicians, Pagni and Maccarrone (respectively Mayor of Pisa and President of the Pisa Province, both members of the CIU board), expressed their regrets: despite the recognition of the potential of an electronic machine, they respectively stated "the synchrotron exerted greater influence on the public opinion" and was "an easy argument for some spectacular propaganda."[22] Moreover, in a January meeting[23] the Faculty of Engineering declared its concerns about the feasibility of the project. Both episodes are examples of the misguided perception the Italian decision makers had about the usefulness of the new calculating devices:[24] in the view of politicians, computers were not marketable to the public, while all things related to atomic energy were a source of fascination towards scientific and industrial progress.

Fermi's letter was read in the October 4 meeting[25] and the choice of the computer was essentially accepted, although some concerns were yet to be overcome and the formal steps to start the project still to be taken. However, only during the January 13-14 meeting[26] the calculator begins to be mentioned as a "suggestion of late prof. Fermi". The scientist died on November 28: the attribution may be a sincere recognition motivated by the emotion for the recent death or, in the wake of the same feelings, a way to overcome a sceptical attitude still lingering inside the University.

In this regard, it is notable the correspondence between the Dean of Engineering Enrico Pistolesi and Avanzi, where the Rector holds a strong position in support of the calculator. In a January 21 letter, Pistolesi argues that a "commission by the Faculty of Engineering should first examine the opportunities and convenience to proceed to the design of the machine, a subject on which many colleagues have expressed concerns".[27] The letter also reveals some of the interests of the Faculty of Engineering: using CIU funding for new buildings. Avanzi's answer on January 31 is firm: he reiterates that, as decided in previous meetings, the control of the project will be entrusted to a University committee and renews to Pistolesi "the request of nominating the representative of his Faculty".[28]

Fermi's support was also used to prevent INAC opposition. The Rome institute was the most advanced research centre in Italy devoted to computing and negotiating the purchase of a Ferranti computer[29]. In a letter to Conversi, Aldo Ghizzetti reports the severe words of the INAC head at the news of the CEP project "I deplore the Pisa initiative of building an electronic calculating machine... I will oppose with all my strength to the waste of money that would occur as a consequence of the approval of the Pisa initiative."[30] In their letters to Picone, along with a diplomatic deference (the CEP is declared as a "definitive second" compared with the INAC Ferranti), Conversi and Bernardini use Fermi as an unquestionable supporter of the project.[31] The story has a happy ending:[32] the position of Picone begins to dissolve in December[33] and INAC will collaborate actively with the CEP project.

On March 9,[34] the CSCE is finally established: the steering committee members are from Physics (Conversi), Mathematics (Alessandro Faedo), and Engineering (Ugo Tiberio). Given the now general agreement on the project, in the foreword the choice of the calculator is simply described as the result of a discussion "at the Congress of Physics held in Varenna with foreign colleagues."

The words "Fermi's suggestion" appear in the minutes once.[35] No information occurs either in the brochures published in 1959[36] or in popular articles[37] appearing in those years. Fermi is not cited during the inauguration of the academic year 1958/59 when the Rector recalls the CSCE along with its first success, the building of the Macchina Ridotta.[38] While the scientist's support was likely decisive for the start of the project, it is a rhetoric overstatement to identify him as *the originator* of the CEP project. The facts could be aptly summarized by quoting a 1958 internal CSCE note:[39] the choice of building a computer was the result of "consultations that professors of the University of Pisa had in Varenna in July 1954 with various internationally renowned physicists, among whom we should remember, in particular, the name of Enrico Fermi". Identifying the CEP project as "Fermi's last gift to Italy"[40] has instead become a topos in the history of Italian computer science, strengthened over time in official ceremonies and even in the memories of the witnesses.[41]

## The very first machine: the Macchina Ridotta

The CSCE Committee[42] had a function of general control, also managing the relations with the academic and scientific world. A research team carried out the study and design activities. The role of Olivetti is cited since the earliest meetings,[43] but the collaboration was formalized in May 1956.[44]

Initially, the leading exponents team were Alfonso Caracciolo, Giuseppe Cecchini, Elio Fabri and Sergio Sibani.[45] Early in the project the Olivetti engineer Mario Tchou was involved with the role of administrative manager[46], but later on he was fully absorbed by the *Laboratorio Ricerche Elettroniche* (LRE, meaning Electronic Research Laboratory) set up by Olivetti in 1955 in Barbaricina, in the Pisa suburb.[47] While Caracciolo and Fabri dealt with the logical and architectural design of the machine, Cecchini and Sibani designed the electronic implementation.[48] The project staff grew and in March 1958 it amounted to over thirty people, including technical and administrative collaborators.[49]

The first year of the CSCE was a period of study with the researchers engaged in acquiring the necessary know-how. As an internal note[50] points out, they initially had some financial difficulties due to delays in funding. Since early 1956, however, the activities were intense with many feasibility experiments.[51] At the end of this period, the CSCE obtained the first important result: the completion of the detailed project of an electronic computer, the "Macchina Ridotta"[52] (Smaller Machine, in the following denoted by the acronym MR). Caracciolo, Cecchini, Fabri and Sibani signed the project; Menotto Baldeschi, Giovan Battista Gerace and Vladimiro Sabbadini were acknowledged.[53]

Most of the surviving technical information on the MR consists of drawings.[54] It is worth to note that these documents refer to a first detailed design, which is substantially different from the machine that was built. Another Report is dated April 1957 and it precedes by a few months the MR completion.[55] It details the numerous changes with respect to the 1956 project, mainly due to the need of increasing the usability and the ability to be connected with multiple input and output devices. The note is authored by the logical-mathematical section of CSCE: Caracciolo and Fabri. In fact, most modifications, although substantial, were of architectural nature and required no changes to the electronic design of the single components. The Report mentions several new blueprints that are no longer present either in the Archives of the University of Pisa or in those of CNR. Only one was recovered, yet an essential one for understanding the MR: the new version of the general schema.[56]

The MR is ready in July 1957, as witnessed by a letter that Conversi addressed to many colleagues of the Italian universities.[57] Besides giving news of the milestone, the scientist discusses the immediate usage of the MR for computing services, making it available to his colleagues and starting to look for real case studies to test the capabilities and the performance of the computer.

Indeed, the manual witnesses the presence of MR users.[58] In 1958 four researchers from four INFN centres joined CSCE, with the aim of acquiring skills on computer programming. The participation of two women, Elisabetta Abate and Marisa Romè, in that group is unusual for the period and it is an exception worth of noting. Abate had the task to write up the user manual of MR: the practical and succint style of the document shows it was thought for users, possibly from outside CSCE, writing algorithms in the machine language.[59] The document lacks any instruction on how to operate the MR: a task for those same designers and engineers who built it.

A careful examination of the recovered blueprints reveals interesting details, like the dating of the drawings, which is sometimes contrary to expectations. Considering for instance the binary adder blueprints, the logical network (the specification[60]) is dated July 11, 1956, while part of the electronic circuit (its implementation) is earlier, June 20 (sum circuit[61]) and June 16 (carry circuit[62]). Moreover, in the two electronic blueprints the notation for the logic gates is reversed: in one case white triangles are AND gates and black ones OR gates, the opposite in the other. Moreover, the electronic circuits have a small flaw that precludes them to operate properly. All these problems were obviously solved, yet they left no trace in the surviving blueprints. Anecdotally, they confirm that, with respect to project documentation, computer scientists' bad habits have deep roots.

Finally, the analysis of the blueprints permits to compare the MR with the technology of the time and to understand the ways in which the Pisa researchers built their know-how. The adder logical network is made according to a solution that reduces the levels of logic gates and, therefore, the calculation time. This solution was published in an article describing the arithmetic unit of the IBM 701.[63] The comparison with the CSCE blueprints leaves no doubt about the sources that inspired the MR designers. The relationship (which is not declared, since neither the article nor at least the IBM origin of the solution are mentioned in the CSCE documents) with the IBM "Defense Calculator" further enriches the story of the CEP project.

## The accomplishments of the Macchina Ridotta

In early 1958, the MR began to be used for research purposes. There are several accounts of the use of the MR outside CSCE. The first calculation service was requested by the Institute of Mineralogy of the University of Pisa for a work on crystallography. The task was completed in April 1958, and the program execution took eighty minutes.[64] Other services include a computation that took about 60 hours, organized in multiple sessions, part of a research on radio frequencies in the ionosphere of the University of Rome. By the end of 1958, the MR had done around 150 hours of outside services for a value that "may be estimated at 8 million liras", as stated in an internal report.[65]

The birth of CSCE had been mentioned in 1956 in the newsletter edited by the Research Centre of the US Navy,[66] but the publication of the first results obtained by the MR aroused greater interest. Right at the end of his tenure, the Rector Avanzi received the request of the scientific attaché of the American Embassy in Rome to visit the CSCE; the meeting took place on October 30, 1959.[67]

Beyond its chronological primacy in Italy, it is worthwhile to compare the MR with the technology of the period. From this point of view, in fact, the MR adopted state of the art solutions that were not easy to find, all at once, on other machines:

- *parallel bit processing*; when the CEP project started most computers were "serial", that is, the bits of a memory word were processed one at a time in successive machine clock cycles; the MR is instead "parallel", that is, capable, like today computers, to process all the bits of a word in a single machine clock cycle;
- *ferrite core memory*; instead of adopting one of the mainstream memory technologies of the early Fifties, like magnetic drums, acoustic delay lines or Williams tubes, the CSCE researchers chose ferrite cores, adopting an emerging technology that will be dominant for two decades;
- *micro-programmed control*; the idea comes from the Cambridge research group, published in 1953 and implemented on EDSAC 2;[68] the MR micro-programming used a less sophisticated technology (removable diodes instead of ferrite rods) and was made easier by the low number of instructions (32), all performed in two micro-instructions (a fetch one and an execute one, without cycles: a pure RISC, we would say), yet the MR is one of the first micro-programmed computers to be operational.

As a confirmation of the commitment of the CSCE research team in trying to develop a state of the art computer, none of the three solutions occurred on the other two computers that in 1955 were present in Italy, the USA CRC102 at the Milan Polytechnic and the British Ferranti Mk1* at INAC. These machines, geographically close and working within public research facilities, could have been studied and used as models by the CSCE researchers. Instead, and not without some risks, new roads were taken that, in little time, produced a working result.

Also with respect to performance, the MR was a fine machine. With careful tuning, the execution time of the instructions was lowered by 30% with respect to the initial estimates, resulting in a clock cycle of 4 or 8 μs depending on the current micro-instruction. The resulting performance of more than 60000 instructions per second was claimed "superior to all existing machines on the market, including the IBM 704 that is located in Paris".[69] It should be noted that the "superiority" stated by Conversi only affects the speed and it is biased by different sets of instructions. The 704 had more memory and peripherals, not to mention the Fortran compiler. However, challenging IBM on the most straightforward benchmark was a result to be proud of.

For the sake of fairness, we must point out an evident flaw in the MR: although it took into account the use of subroutines, it had no specific jump-to-subroutine instruction. At the time, the *Wheeler Jump* used in the EDSAC was a well-known solution.[70] It might have been implemented on the MR with minimal changes on the hardware. The Pisa researchers did not, although in other respects (the micro-programming) they had adopted results of the Cambridge group.

It is surprising that until the 2011 Pisa conference[71] the technological relevance of the MR had not been properly recognized.[72] A few different causes objectively contributed to such an overlook:

- among the four MR designers, only Caracciolo will remain at CSCE, and only until the early Seventies; Fabri and Sibani left the CSCE in 1959,[73] Fabri going back to astrophysics, Sibani returning to Olivetti, as did Cecchini in 1961;
- there was no surviving physical evidence for the MR, since it was completely dismantled for reusing the electronic materials in the building of the CEP;
- there was no official opening of the MR, due to political tensions between the University of Pisa and the Government, the traditional ceremony for the inauguration of the academic year 1957/58 was not held;[74] the MR will be remembered only in the prolusion of the following year,[75] when it had already been dismantled;
- the CSCE working plan for the years 1956/57 has no mention of a first computer, but of a "machine core, i.e. the entire machine with the exclusion of the external devices: the magnetic drum and the fast input/output system".[76]

The most crucial reason is likely the description of the MR as the "core" of the later CEP, a description that will appear in several official documents later on. Interpreting the MR as an incomplete part of the CEP, it was natural for historians to underestimate its importance.

In fact, after a careful analysis of the technical documents, it is fair to say that the two machines were very different (see also Table 1). For example, the memory, one of the components that would have been easier to reuse with few changes, was completely redesigned: it was made by single-sided 32x32 planes in the first computer and by double-sided 64x64 planes in the second. And this is just one example: from the micro-programmed control to the electronics of the adder, the differences between the two machines are several and significant.

| | Macchina Ridotta, 1957-58 | CEP, 1961-69 |
|---|---|---|
| word | 18 bit | 36 bit |
| memory | 1024 words<br>ferrite core<br>32x32 single-sided planes | 4096 words (later 8192)<br>ferrite core<br>64x64 double-sided planes |
| logic implementation | diode-resistor logic, triode inverters | diode-resistor logic, triode inverters<br>transistor inverters in few cases |
| microprogrammed control | diode matrix<br>fetch or execute microinstructions | ferrite rods matrix<br>pseudo instructions, conditional<br>and cyclic microinstructions |
| number of instructions | 32 | 128 |
| fixed point additions per sec | 62500 | 67000 |
| floating point additions per sec | not available as machine instruction | 10400 |
| support for subroutines and array operations | address substitution | double indirection using index cells |
| I/O devices | 1  teletypewriter Olivetti T2CN<br>1  teletypewriter Olivetti T2CN-PF with tape puncher<br>1  tape reader Olivetti T2TA10<br>1  fast tape reader Ferranti TR5 | 1  teletypewriter Olivetti T2CN<br>2  fast tape readers Ferranti TR6<br>2  fast tape punchers Teletype LMU6<br>1  line printer Bull<br>1  magnetic drum (32768 36 bit words)<br>6  magnetic tape readers (later) |
| system software | basic arithmetic subroutines<br>simple program loader | 220 math and utility subroutines,<br>symbolic assembler,<br>FORTRAN compiler (later) |

Table 1. Technical data of the two computers built by the University of Pisa

Some authors cited the MR with more details,[77] but rather than the machine that was actually built in 1957 they describe the 1956 design, as there are more surviving copies of the corresponding report,[78] but this is the first draft, and it describes a different and simpler computer. By comparing the two designs it is possible to appreciate the differences: the addition of other devices (a second teletypewriter, the tape puncher, a second tape reader), the flexible management of the I/O operations, the easy boot of the system software with a sort of "direct memory access", and the mechanism of hot breakpoints for debugging. Other improvements in terms of usability are reflected in the design of the manual control panel and in the visual feedback on the state of the memory and on the value of the program counter. The comparison between the two versions of the MR design witnesses the remarkable work done by the CSCE researchers between 1956 and 1957. It is a process that shows the importance of MR in the CEP project and, in general, in the formation of the first Italian computer scientists.

Unfortunately, very little photographic documentation of the MR survives. Most telling is the picture in Fig. 1, presenting the three racks that made up the machine: from left to right, the arithmetic-logic unit, the micro-programmed control, and the memory. In the background, under the window you can see the back of the control panel. On the far right of the picture, the second teletypewriter appears on the foreground, equipped with the tape puncher.

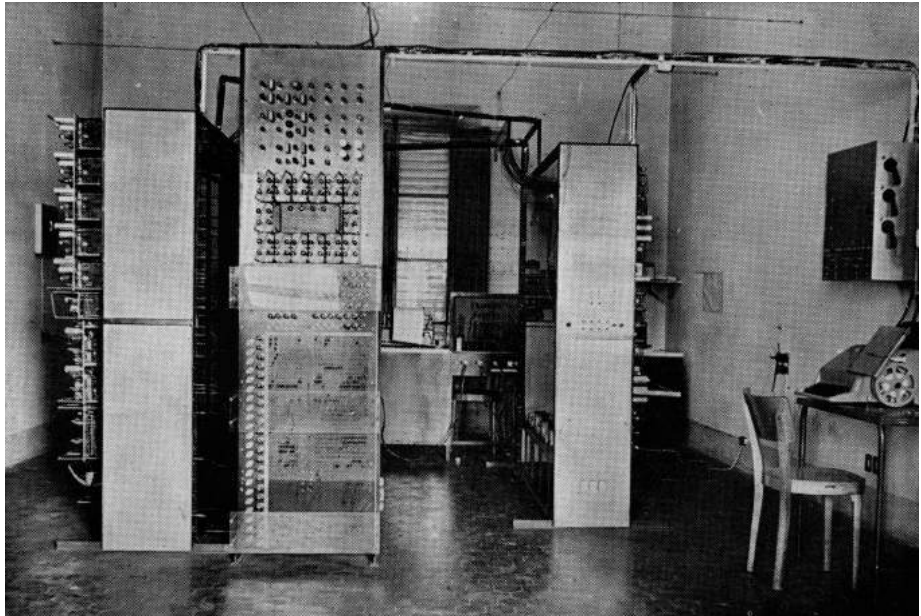

Figure 1. The Smaller Machine

## The difficult years

The MR success closed the first phase of the project. Unfortunately, the next goal, that is the design and building of the CEP, run up in a difficult period for the CSCE.

A first problem was represented by the transistor technology, whose production boomed since mid-1958.[79] More important than these changes, the original financing was nearing the end. The hopes of finding new funds were rosy and lasted for some time. Still in the CIU meeting of April 1960, it was stated that "while the Government will finance the other universities to buy electronic computers, Pisa is the only one that has built a large electronic computer without asking anything to the State".[80] This refers to universities such as Turin, Padua, and Naples, and it was to encourage the attempts by Conversi at the Ministry and at the National Research Council.

Sadly, the achievements of Pisa were not rewarded. Even if the CSCE had proved its ability to keep the pace of the most advanced international projects, it was unable to obtain more funding. The conversion to transistors was thus not possible. Yet, in order to give a display of technological capacity, they were used in the CEP micro-programmed control, which instead of diodes used ferrite rods to implement the ROM matrix (in a variation of the idea proposed by the British EDSAC2).

Besides the additional costs of the hardware, the problems of software development were added. A substantial commitment to the realization of software libraries was early planned.[81] However, still in February 1961, while the machine hardware was missing only a few tests, there were worries about the delay in the development of the system software. In order to work on the CEP software before the completion of its hardware, collaboration was set with the INAC to develop a CEP simulator using the Ferranti computer owned by the Institute, but also this solution was late.[82] The CEP was eventually up and running in late spring 1961.[83]

Despite its limits, the CEP was a relevant machine. US observers[84] recognized its features, including the speed, the micro-programming, and the indirection mechanisms to support subroutines. Some of them, as micro-programming and performance, were already present on the MR: as Blachman bluntly points out, the CEP was late.

The importance of consistent and continuative investment in research is also evident by comparing the CSCE results with those of its partner, Olivetti. The opportunity is offered by a curious telegram sent on December 23, 1960. The CSCE researchers, along with the Christmas greetings, announce to Faedo (by then the new Rector) that the CEP is "working in the majority of its instructions".[85] This self-congratulatory telegram led some to anticipate at December 1960 the completion of the machine.[86] A more careful look reveals instead a bitter ending: as relevant as the text itself is in fact the advertising in the telegram card, where Olivetti was promoting its electronic computers, additionally using the modern term "calcolatore" instead of "calcolatrice" (that is, computer instead of calculator).

While CSCE was suffering for lack of funding, Olivetti invested and accelerated the pace. The LRE in Pisa, almost at the same time as the MR, developed the *Macchina Zero*, the first Olivetti computer (sometimes also named *Elea 9001*). A second prototype, named *Elea 9002*, was completed in 1959, with the electronic redesigned using transistors; it was installed at the headquarters Olivetti in Milan and inaugurated on November 8, 1959. The commercial product, named *Elea 9003*, was announced at the Milan Fair in April 1959 and the first machines were delivered to customers in 1960.[87]

## The CEP vernissage

The CEP was inaugurated on November 13, 1961 at the presence of the Italian President Giovanni Gronchi. Indeed, as stated in a letter from Conversi to Faedo related to an open project call,[88] they tried to anticipate the ceremony to give "a little resonance in time for obtaining some additional funding." The agenda of the President, however, delayed the ceremony to the fall.[89]

The inauguration ceremony[90] was a big event for the city of Pisa. The University press release[91] and the following articles on the newspapers[92] trace the history of an exciting project that saw the enterprising financing of local authorities, the support of a great scientist like Fermi, young researchers from various parts of Italy, the participation of an industrial visionary like Olivetti.

In the memories of the protagonists, though, the inauguration represented the start of a difficult period: the goal was achieved yet the funding ended and the future was uncertain. In December 1961 a new statute was drafted for the CSCE, with the hope of receiving the approval by the National Research Council. In July 1962 an agreement was signed that began the process of transition: the CSCE becomes a "CNR Institute at the University of Pisa".[93] The CNR is now responsible for the staff, while the University is in charge for the facilities. The management is entrusted to a body that sees an equal share of representatives by the University and by the CNR. The convention lasted until 1968 when the CSCE became the CNR Institute for Information Processing.[94]

In the research group there will be rotations and new arrivals, but the Centre thus reaches the stability that will allow the further development of the CEP (e.g. doubling the memory and adding magnetic tape drives) and to use it both for doing research on the software side (e.g. the Fortran compiler) and for offering computing services outside. The CEP will remain in operation until the end of the Sixties, working day and night. Still in 1966, the machine is accounted for an average of more than three hundred machine hours per month, with peaks of almost five hundred: as in March when the CEP worked for 496 hours, 170 of which for users outside the CSCE.[95]

## Concluding remarks

The research summarized in the paper investigated the Pisa archives with the aim of throwing further light on some facets of the CEP project, a seminal enterprise for the development of Computer Science in Italy. Besides the further layers of complexities added to the history of the project, each one of these facets is in itself a litmus test for a different aspect of the research business and practice.

Removing the almost hagiographical aspects, the involvement of Enrico Fermi in the beginning of the CEP project helps in delineating the real issues at stake and the role of the institutional players, as well as the need for public endorsement and knowledge dissemination.

The exploration of the MR accomplishments witnesses the importance of moving beyond the documents into a careful analysis of the technical issues, in order to faithfully reconstruct the results of a research endeavour and its connections with other projects around the world, thus properly assessing its merits. Also, the analysis can be exploited for didactic purposes, as witnessed by the use of the original blueprints to build the MR simulators[96] and the working replica of the 6-bit adder[97].

Finally, the difficulties in the building the 1961 CEP is a *cas exemplaire* on the need of continuous funding for the technological research, either by the public hand or by the private sector. A lesson that is still valid today as it was in the Fifties.

## Nota bene: References to documents

The [n] item added to the archive documents is a reference to the list present at the webpage http://hmr.di.unipi.it/IEEEAnnals, where digital copies of such documents have been made available for the sake of the reader. Often not catalogued or of difficult accessibility, these documents are from the General Archive of the University of Pisa (*AUniPi*), the Library of the Institute of Science and Technology of Information, CNR Pisa (*AISTI*), the Archive of the Institute for Applications of Computing, CNR Roma (*AIAC*), and the private archive of Elio Fabri (*AFabri*).